\documentclass[aps,prl,superscriptaddress,reprint,floatfix]{revtex4-1}
\usepackage[utf8]{inputenc}
\usepackage{amsmath}
\usepackage{amssymb}
\usepackage{graphicx}
\usepackage[colorlinks,urlcolor=blue]{hyperref}
\usepackage{url}
\usepackage{color,xcolor}
\usepackage{ulem}
\usepackage{float}
\usepackage{epstopdf}
\usepackage[none]{hyphenat}
\begin{document}

\title{Compressive Strain Turns $s^{\pm}$ into $d$-Wave Pairing in One-unit-cell La$_3$Ni$_2$O$_7$ Thin Film Via Substrate-Induced Hole Doping}
\author{Yang Zhang}
\affiliation{Materials Science and Technology Division, Oak Ridge National Laboratory, Oak Ridge, Tennessee 37831, USA}
\author{Ling-Fang Lin}
\email{lflin@utk.edu}
\affiliation{Department of Physics and Astronomy, University of Tennessee, Knoxville, Tennessee 37996, USA}
\author{Adriana Moreo}
\affiliation{Department of Physics and Astronomy, University of Tennessee, Knoxville, Tennessee 37996, USA}
\affiliation{Materials Science and Technology Division, Oak Ridge National Laboratory, Oak Ridge, Tennessee 37831, USA}
\author{Satoshi Okamoto}
\affiliation{Materials Science and Technology Division, Oak Ridge National Laboratory, Oak Ridge, Tennessee 37831, USA}
\author{Thomas A. Maier}
\email{maierta@ornl.gov}
\affiliation{Computational Sciences and Engineering Division, Oak Ridge National Laboratory, Oak Ridge, Tennessee 37831, USA}
\author{Elbio Dagotto}
\email{edagotto@utk.edu}
\affiliation{Department of Physics and Astronomy, University of Tennessee, Knoxville, Tennessee 37996, USA}
\affiliation{Materials Science and Technology Division, Oak Ridge National Laboratory, Oak Ridge, Tennessee 37831, USA}

\date{\today}

\begin{abstract}
Motivated by recent reports of ambient-pressure superconductivity in La$_3$Ni$_2$O$_7$ films grown on LaSrAlO$_4$, we investigate the superconducting instability in a one-unit cell thin film using {\it ab initio} and random-phase approximation techniques. Compared to the high-pressure bulk system, the ratio of inter-layer $d_{3z^2-r^2}$ hopping to intra-layer $d_{x^2-y^2}$ hopping is suppressed in the 1UC thin film, and the crystal-field splitting of the $e_g$ orbitals is increased. Our calculation indicates that spin-fluctuation-driven pairing correlations are weak for the stoichiometric case at ambient pressure, but increase significantly under hole doping. The leading pairing symmetry is also found to change by hole doping. Specifically, we obtain a leading $d_{x^2-y^2}$ pairing state at moderate hole doping, followed by a $d_{xy}$ state at higher doping. These states are driven by intra-band spin-fluctuation scattering {\it within} the $\gamma$ hole pocket centered around the M point, and arises primarily from states in the Ni layer {\it farther} from the substrate. These results strongly suggest that the thin-film superconducting samples are hole-doped and that pairing in this system predominantly arises in the layer,
as opposed to the inter-layer pairing in the pressurized bulk system.
\end{abstract}

\maketitle
 {\it Introduction--} The exciting discoveries of high-$T_c$ superconductivity in Ruddlesden-Popper (RP) nickelates bilayer (BL) and trilayer systems under pressure~\cite{Sun:arxiv,Zhu:arxiv11,Li:cpl,Zhang:arxiv-exp,Hou:arxiv,Sakakibara:arxiv09,Wang:arxiv9,Wang:nature,Zhang:arxiv11} have stimulated intense interest in the field~\cite{Zhang:prb23,LiuZhe:arxiv,Yang:arxiv09,Zhang:prb24,Dong:arxiv12,Xie:SB,Chen:arxiv2024,Dan:arxiv2024,Zhang:prb23-2,Chen:prl24,Sakakibara:prl24,Lin:prb23,Oh:prb24} and expanded the exploration of nickelate superconductivity beyond the infinite-layer family~\cite{Li:Nature,Zhang:prb20,Nomura:rpp,Pan:nm}, opening a remarkable platform for unconventional superconductivity~\cite{Chen:prb24,Ryee:prl,Zhang:prl, LaBollita:prm,Lu:prb25,Duan:arxiv25,Braz:arxiv25,Luo:prl23,Zhang:arxiv24,LaBollita:prb24,Qu:prl,Lu:prl,Tian:prb24,Liu:arxiv2023,Takegami:prb,Xu:prb25,Zhang:2323,Shi:arxiv25,Zhang:1212,Zhang:1313,Zhang:nc24,Christiansson:arxiv,Yang:arxiv,Liu:arxiv,Yang:arxiv1,Oh:arxiv,Liao:arxiv,Lechermann:arxiv,Fan:STM,Shilenko:arxiv,Huang:arxiv,Qin:arxiv,Wang:PRB24,Cao:arxiv,Maier:arxiv25,Li:scpma}.

 While experimental efforts continue to improve the sample quality, transition temperature $T_c$, and superconducting volume fraction under pressure~\cite{Li:arxiv25,Lip:nsr25,QiuZ:arxiv25,Guo:arxiv25,Dong:arxiv25}, recently a remarkable discovery was made, i.e., ambient-pressure superconductivity in thin-film BL nickelates grown on compressively strained ($-2\%$ strain) LaSrAlO$_4$ (LSAO)~\cite{Ko:nature}, extending investigations that were previously limited to high-pressure bulk systems~\cite{Liu:nm25,Zhou:nature,Geisler:arxiv25,Huang:arxiv25,Hao:arxiv25,Ushio:arxiv25,Qiu:arxiv25,Cao:arxiv25,Shao:arxiv25,Xiang:arxiv25,Ji:arxiv25,Lv:arxiv25,Bleys:arxiv25,Liu:arxiv25,Osada:cp25,Xiong:arxiv25,Tarn:arxiv25}. The original thin-film experiments on La$_3$Ni$_2$O$_7$~\cite{Ko:nature} report an onset $T_c$ ranging from 26 to 42 K. By partial Pr$^{3+}$ replacement and optimization of growth conditions, the onset and zero resistance $T_c$ are increased to 48 K and 30 K~\cite{Liu:nm25}, respectively, while the Meissner diamagnetism is also observed in (La,Pr)$_3$Ni$_2$O$_7$ thin films~\cite{Zhou:nature}.

These breakthroughs in thin-film superconductivity offer an excellent opportunity to investigate the electronic structure of the superconducting state in BL nickelates. However, recent angle-resolved photoemission spectroscopy (ARPES) measurements remain controversial: while Refs.~\cite {Shen:arxiv25,Li:nsr25} found a small $\gamma$ pocket originating from the $d_{3z^2-r^2}$ orbitals bonding state, Ref.~\cite{Wang:arxiv25} reported that the $\gamma$ band does not cross the Fermi level.

\begin{figure}
\centering
\includegraphics[width=0.42\textwidth]{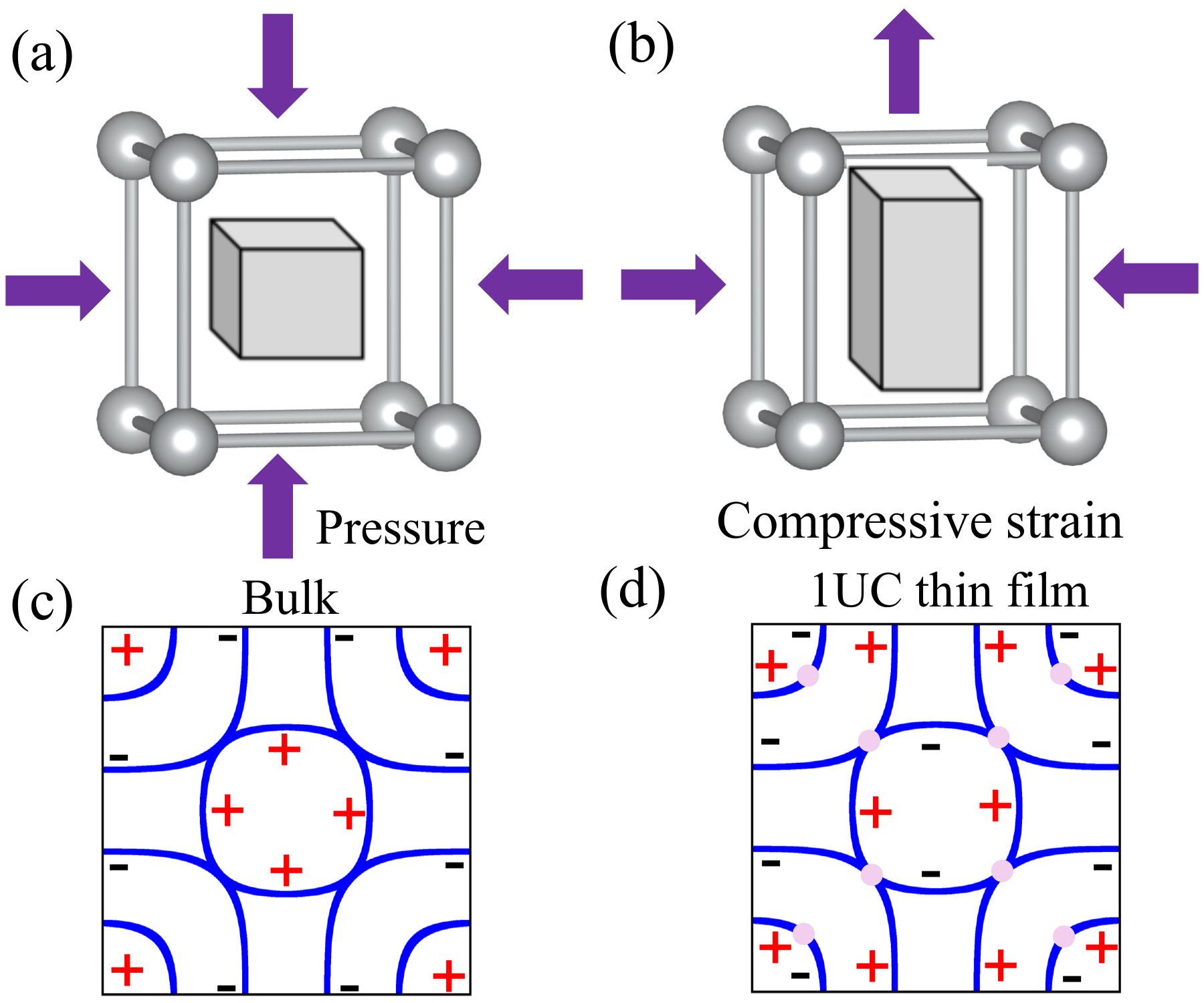}
\caption{ (a) Schematic of lattice modifications under the high-pressure and in-plane compressive-strain thin-film conditions. (c-d) Sketches of the Fermi surfaces for the high-pressure and compressive strain for hole-doping thin-films, including the signs of the superconducting order parameter. (c) is $s^{\pm}$ and (d) is $d_{x^2-y^2}$. The light pink dots in (d) denote nodes.}
\label{Fig1}
\end{figure}

Here, using density functional theory (DFT)~\cite{Kresse:Prb,Kresse:Prb96,Blochl:Prb,Perdew:Prl,Liechtenstein:prb,Mostofi:cpc} combined with random phase approximation (RPA)~\cite{Graser2009,Kubo2007,Romer2020,Altmeyer2016,Mishra:sr,Maier:prb22} calculations, we found similarities but also key differences between compressively strained one-unit-cell (1UC) thin film and high-pressure BL nickelates. The inter-layer hopping between two Ni layers is considerably reduced in the thin-film case, leading to weakening of the dimerization of the $d_{3z^2-r^2}$ orbitals. Furthermore, the splitting of the crystal field between the two $e_g$ orbitals is enhanced in the thin-film format. For the stoichiometric 1UC case ($n = 3.0$), our RPA calculations reveal only a very weak singlet pairing strength. However, we found that a leading spin-fluctuation driven $d$-wave pairing instability increases significantly upon hole doping. Comparing the corresponding Fermi-surface shape with recent ARPES results for the 1UC sample~\cite{Li:nsr25}, our theoretical results suggest that the superconducting BL in the thin-film sample is indeed hole-doped. While the Fermi surface topology is similar in both cases, the superconductivity is driven by intra-layer singlet pairs in the thin-film, resulting in a $d$-wave state. This differs from the inter-layer singlet-pair-driven $s^{\pm}$-wave instability observed in the high-pressure bulk systems, as illustrated in Fig.~\ref{Fig1}.

\begin{figure}
\centering
\includegraphics[width=0.44\textwidth]{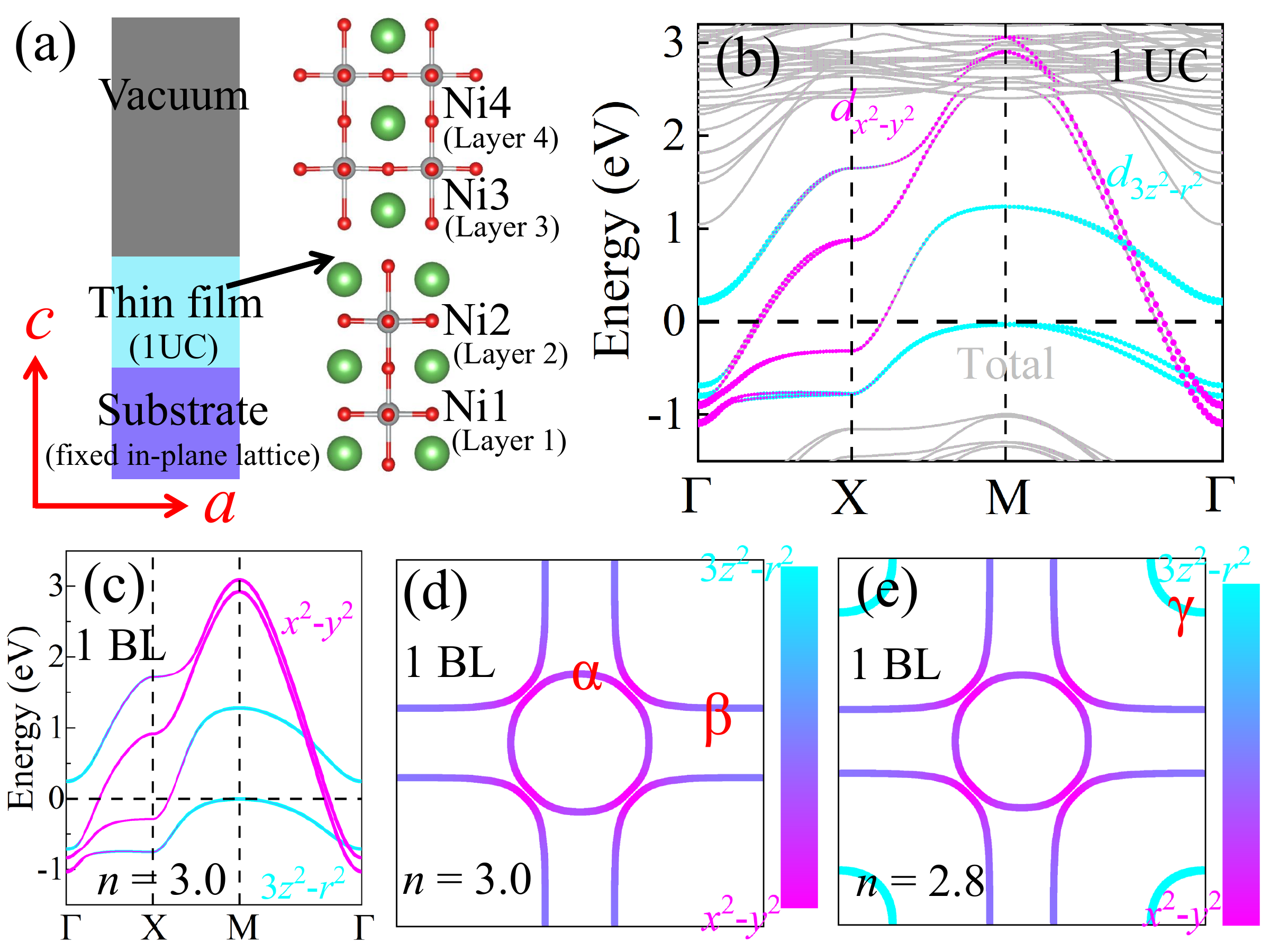}
\caption{ (a) Schematic structural slab model and crystal structure (green = La; gray =Ni; red = O) of the 1UC thin film, with two BL blocks (four Ni layers) visualized via the VESTA code~\cite{Momma:vesta}. Note that the local $z$-axis is perpendicular to the NiO$_6$ plane toward the top O atom, while the local $x$ or $y$ axis is along the in-plane Ni-O bond directions. The in-plane lattice constants are fixed to those of the LSAO substrate ($a = b = 3.7557$~\AA), and a vacuum layer of more than 20.0~{\AA} is used to simulate the thin-film slab geometry. (b) Projected band structures for the 1UC thin film. The electronic structures are calculated within the local density approximation plus Hubbard $U$ and Hund’s coupling $J$ with the Liechtenstein formulation for the double-counting term~\cite{Liechtenstein:prb}. Here we use $U = 3.8$ eV and $J = 0.6$ eV, values obtained from the constrained random-phase approximation for La$_3$Ni$_2$O$_7$~\cite{Christiansson:arxiv}. (c-e) Tight-binding band structure and Fermi surface for the one BL (two Ni layers) model for the thin-film at (c-d) with $n = 3$, and (e) with $n = 2.8$, respectively. The hopping file can be found in the SM~\cite{Supplemental}.}
\label{Fig2}
\end{figure}

{\it Compressive strain with the LSAO substrate--}  To simulate the strain effect induced by the substrate, we consider the slab model [see Fig.~\ref{Fig2}(a)] and optimize the atomic positions, while keeping the in-plane lattice constant fixed to that of the LSAO substrate. As shown in Fig.~\ref{Fig2}(b), the bonding state of $d_{3z^2-r^2}$ orbitals does not touch the Fermi surface ($\sim 10$ meV below the Fermi level), and therefore the $\gamma$ pocket around the M point is absent, in contrast to the high-pressure bulk systems. Compared to the optimized high-pressure bulk~\cite{Zhang:nc24,Zhang:prb23}, the inter-layer hopping between $d_{3z^2-r^2}$ orbitals is decreased by approximately $15\%$, leading to a reduced bonding-antibonding splitting energy in the thin film. By contrast, the intra-layer hopping between $d_{x^2-y^2}$ orbitals remains almost unchanged. Thus, the ratio $t^z_{\alpha\alpha}/t^{x/y}_{\beta\beta}$ ($\alpha$: $d_{3z^2-r^2}$ and $\beta$: $d_{x^2-y^2}$) of inter-layer hopping to intra-layer hopping is suppressed to $\sim$1.19 in the thin film, compared to that in the high-pressure bulk ($\sim 1.31$)~\cite{Zhang:nc24}. Moreover, due to the symmetry-breaking geometry of 1UC, Ni1 (or Ni4) and Ni2 (or Ni3) are asymmetric (see Fig. 2(a)): we observed that the $d_{3z^2-r^2}$ orbital at the Ni2 site always has more electrons than that in the Ni1 site at the same overall fillings. For example, for 1BL, the $d_{3z^2-r^2}$ population occupation on Ni2 is $\sim 1.16$ while $\sim 0.85$ for Ni1 (See details in Supplemental Materials (SM)~\cite{Supplemental}).  This originates from the increased Ni2 apical O distance ($\sim 2.36$ \AA) compared with the Ni1 apical O distance towards the substrate ($\sim 2.15$ \AA), because expanding into a vacuum is energetically easier. This lowers the $d_{3z^2-r^2}$ orbital level in the top NiO$_2$ plane and facilitates its electron occupation.

Next, based on the hoppings obtained from the 1UC case~\cite{Supplemental}, we construct a four-band $e_g$-orbital tight-binding model on the one BL lattice for the thin-film case with overall filling $n = 3$, as shown in Figs.~\ref{Fig2}(c,d) (details are shown in Sl of SM~\cite{Supplemental}). The calculated average electronic density of the $d_{3z^2-r^2}$ orbitals (1.01 per Ni site) is slightly higher than that of high-pressure bulk (0.814 per site)~\cite{Zhang:arxiv24}, for the stoichiometric ($n = 3$) filling in the tight-binding model. By comparing the areas of the $\gamma$ pocket vs. the ARPES experiments for the 1UC case~\cite{Li:nsr25}, our tight-binding Fermi surfaces indicates that the density $n = 2.8$ (see Figs.~\ref{Fig2}(e)) in our model is realistic, suggesting a hole-doped scenario for the superconducting thin films, probably caused by Sr migration from substrate to sample.

\begin{figure*}
\centering
\includegraphics[width=0.88\textwidth]{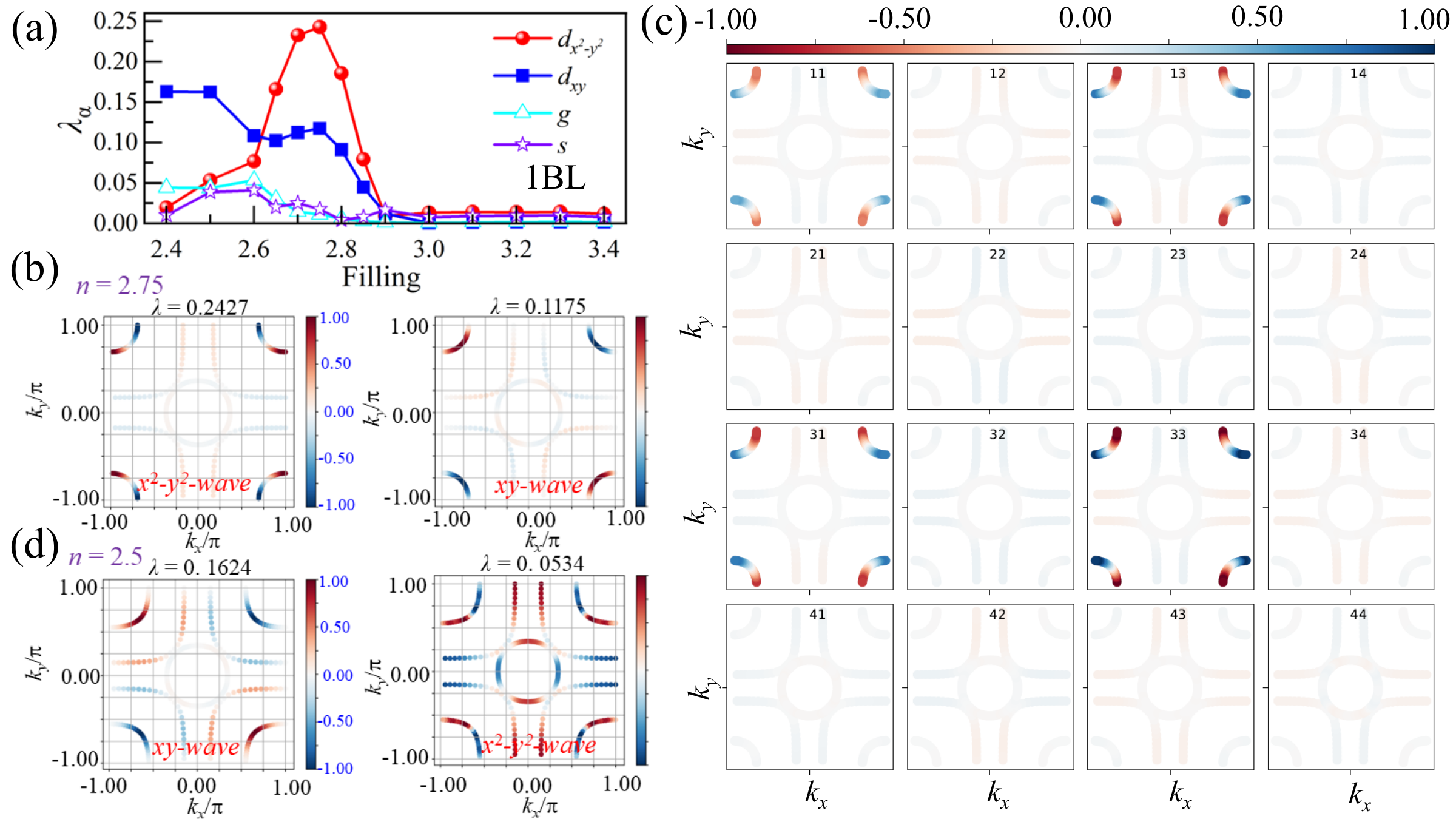}
\caption{ (a) The RPA calculated pairing strength $\lambda$ for different instability channels vs. electron density fillings in the compressive-strain BL model. (b,d) The RPA calculated leading superconducting singlet gap structures $g_\alpha({\bf k})$ for momenta ${\bf k}$ on the Fermi surfaces for $d_{x^2-y^2}$-wave and $d_{xy}$-wave with corresponding pairing strengths $\lambda$ for the case of the LSAO substrate at (b) $n = 2.75$ and (d) $n = 2.5$, respectively. (c)  Gap structure in orbital space for $n = 2.75$ with the indexing: 1 = $d_{3z^2-r^2}$ orbitals from layer 1 (Ni1), 2 = $d_{x^2-y^2}$ orbitals from layer 1 (Ni1), 3 = $d_{3z^2-r^2}$ orbitals from layer 2 (Ni2), and 4 = $d_{x^2-y^2}$ orbitals from layer 2 (Ni2). The sign of $g_\alpha({\bf k})$ is indicated by colors (red = positive, blue = negative), and its amplitude by the color intensity. Here, we used Coulomb parameters $U=0.7$ eV~\cite{RPA-U}, with the ratios $U'=U/2$, and $J=J'=U/4$ already employed in previous literature~\cite{Zhang:nc24,Zhang:prb23-2}. The calculation was performed at $T = 0.01$ eV.}
\label{1uc}
\end{figure*}

{\it Pairing tendency in thin films--} To explore the superconducting pairing tendencies of the 1UC thin film nickelate, we carry out multi-orbital RPA calculations to assess the BL tight-binding model, including on-site Coulomb repulsion terms, i.e., the intra-orbital $U$, inter-orbital $U'$, Hund's rule exchange $J$, and pair hopping $J'$ terms. RPA is based on a perturbative weak-coupling expansion in these Coulomb interaction terms~\cite{Kubo2007,Graser2009,Altmeyer2016,Romer2020}. The pairing strength $\lambda_\alpha$ and the gap structure $g_\alpha({\bf k})$ for channel $\alpha$ are obtained by solving an eigenvalue problem of the form
\begin{eqnarray}\label{eq:pp}
	\int_{FS} d{\bf k'} \, \Gamma({\bf k -  k'}) g_\alpha({\bf k'}) = \lambda_\alpha g_\alpha({\bf k})\,.
\end{eqnarray}
The momenta ${\bf k}$ and ${\bf k'}$ are restricted to the Fermi surface, and the singlet pairing interaction $\Gamma({\bf k, k'})$ is given by the irreducible particle-particle vertex. Within RPA, the dominant term entering $\Gamma({\bf k, k'})$ is the RPA spin susceptibility $\chi({\bf k-k'})$.

Figure~\ref{1uc} shows the leading eigenvalues (pairing strengths) and corresponding eigenvectors (gap structures) found by solving the eigenvalue problem in Eq.~(\ref{eq:pp}). As shown in Fig.~\ref{1uc}(a), the pairing strength $\lambda_\alpha$ is negligibly small for the stoichiometric case with filling $n=3$ for all the singlet channels $\alpha$. This pairing strength, however, increases drastically when the system is doped with holes ($n < 3$), reflecting the sharp peak in the density of states at the top of the $\gamma$-band.

For $n\lesssim 2.85$, we find that two different $d$-wave solutions develop significant pairing strengths. This includes a leading $d_{x^2-y^2}$ solution at moderate hole doping with a dome-like doping dependency, and a $d_{xy}$ solution, which becomes leading at higher doping for $n\lesssim 2.6$. As displayed in Fig.~\ref{1uc}(b) for $n=2.75$, the $d_{x^2-y^2}$ state is almost exclusively restricted to the $\gamma$ pockets at the M point, as is the sub-leading $d_{xy}$ state at this doping. In Fig.~\ref{1uc}(c), to analyze the orbital characters of the leading $d_{x^2-y^2}$ state, we show the gap structure in the orbital space defined as $a_{\ell_1\ell_2}({\bf k}) = \sum_\nu g_{\nu}({\bf k}) a_\nu^{\ell_1}(\bf k)a_\nu^{\ell_2}(-\bf k)$, where $a_\nu^{\ell}({\bf k})$ are the band ($\nu$)-orbital ($\ell$) matrix elements that diagonalize the non-interacting tight-binding Hamiltonian. This leading $d_{x^2-y^2}$ state is characterized by purely intra-orbital contributions from the $d_{3z^2-r^2}$ orbital (see layer 1 for index 1, layer 2 for index 3). The dominant contribution corresponds to the intra-layer (33) term from the Ni2 layer (the layer farther from the substrate). In addition, the inter-layer term (13), followed by the intra-layer (11) term from the Ni1 layer, also provides secondary contributions showing that all the quantitatively relevant contributions arise from orbital $d_{3z^2-r^2}$. In contrast, the contributions from the $d_{x^{2}-y^{2}}$ orbital are negligible.

As the hole concentration increases further, the pairing strength of the $d_{x^2-y^2}$ state sharply decreases, and the $d_{xy}$ state becomes the leading solution for $n \lesssim 2.6$ (see Fig.~\ref{1uc} (a)). The momentum structure of this $d_{xy}$ state, as well as the $d_{x^2-y^2}$ state for $n=2.5$, is shown in Fig.~\ref{1uc} (d). While the amplitude is still largest on the $\gamma$ pocket, it now also has a significant contribution from the $\beta$ sheet, which is in-phase (out-of-phase) with that on the $\gamma$ sheet for the  $d_{xy}$ ($d_{x^2-y^2}$) state.

For electron doping $n>3$, by contrast, the pairing strength is negligible for all the channels we have studied. These results suggest that thin film systems, for which superconductivity has been observed, are hole-doped. This is in agreement with ARPES results~\cite{Li:nsr25} and may be related to a hole-transfer from the substrate to the nickelate BL, or due to oxygen vacancies.

\begin{figure}
\centering
\includegraphics[width=0.44\textwidth]{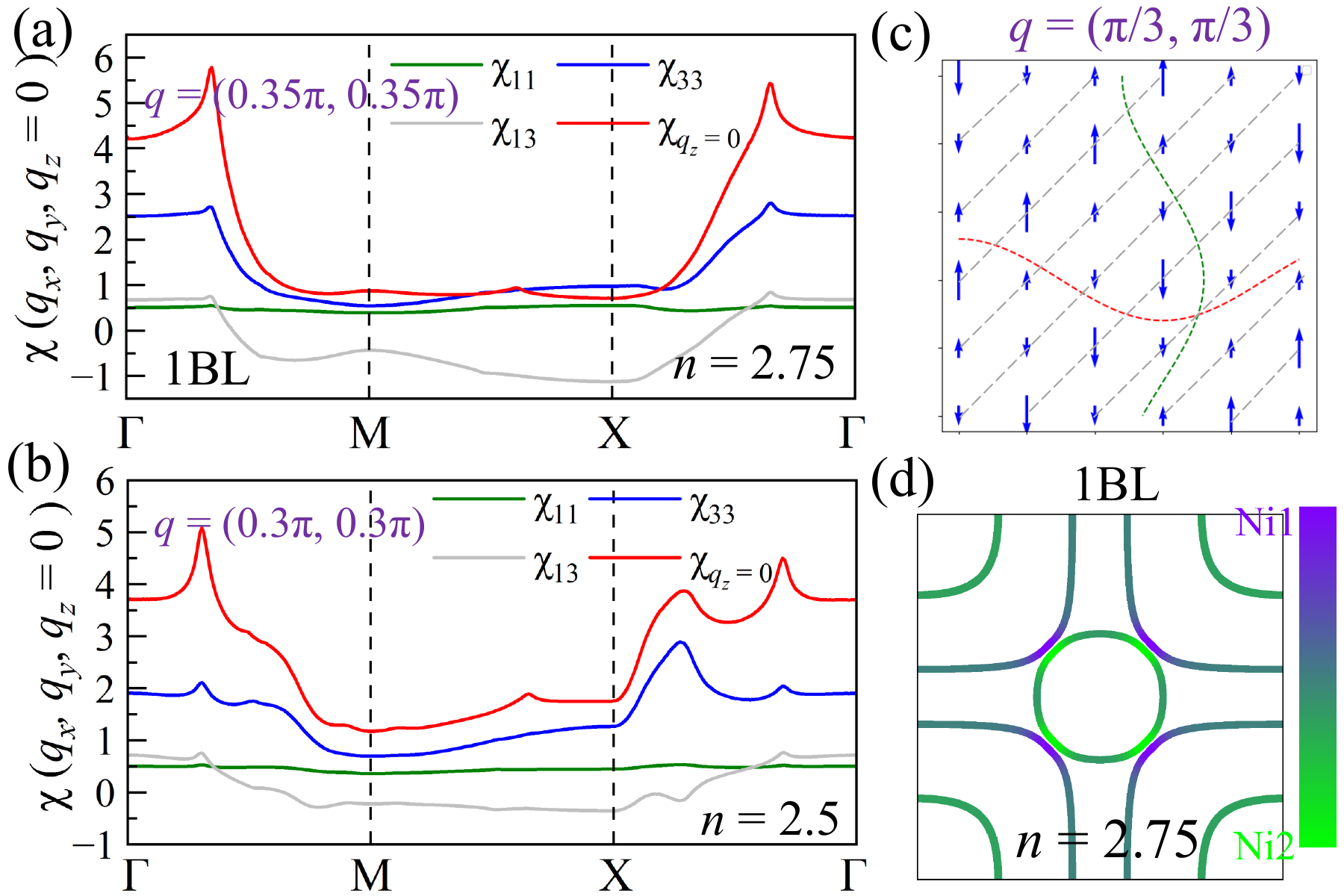}
\caption{(a-b) The RPA calculated static spin susceptibility $\chi'({\bf q}, \omega=0)$ vs. $q_x$ and $q_y$  for the two-orbital 1BL model for $n=2.75$ and $n = 2.5$ respectively, and its orbital and layer contributions.  (c) Spin patterns in real space for ${\bf q}$ = $(\pi/3, \pi/3)$ using a $6\times6$ cluster ($\pi/3$ is used instead of 0.35 for proper fit in the cluster), while a more detailed analysis for  ${\bf q}$ = $(0.35\pi, 0.35\pi)$ and  ${\bf q}$ = $(0.3\pi, 0.3\pi)$ is in the SM~\cite{Supplemental}. Dashed lines represent the waveforms along the two directions. (d)  The layer contributions (Ni1: NiO$_2$ layer 1, Ni2: layer 2) to the Fermi-surface Bloch states for $n = 2.75$.}
\label{Momentum-structure}
\end{figure}

To understand the origin of the leading $d_{x^2-y^2}$-wave and $d_{xy}$-wave pairing instabilities, we now examine the structure of the RPA spin susceptibility tensor $\chi({\bf q})$ at $n = 2.75$ and $n = 2.5$, which is obtained from the Lindhard function $\chi_0({\bf q})$ as
\begin{eqnarray}
	\chi({\bf q}) = \chi_0({\bf q})[1-{\cal U}\chi_0({\bf q})]^{-1}.
\end{eqnarray}
Here, all the quantities are rank-four tensors in the orbital indices $\ell_1, \ell_2, \ell_3, \ell_4$ and ${\cal U}$ is a tensor involving the interaction parameters~\cite{Graser2009}.
The physical spin susceptibility is obtained by summing the pairwise diagonal parts of the tensor, $\chi_{\ell_1\ell_1\ell_2\ell_2}({\bf q})$, over $\ell_1$, $\ell_2$.

Figure~\ref{Momentum-structure} shows the static RPA spin susceptibility $\chi'({\bf q}, \omega=0)$ vs. in-plane ${\bf q}$=$(q_x, q_y)$ for $n = 2.75$ and $n = 2.5$ in panels (a) and (b), respectively, along with their leading orbital contributions. This data is for the even combination of the two layers, which captures the out-of-plane ferromagnetic ($q_z=0$) correlations. We find that for both of these fillings, the $q_z=0$ susceptibility is stronger than the  $q_z=\pi$ (odd layer combination) susceptibility, which represents out-of-plane antiferromagnetic correlations. This is in stark contrast to the pressurized bulk system, for which we previously found that the odd $q_z=\pi$ antiferromagnetic fluctuations dominate, scattering between the bonding band (that creates the $\gamma$ Fermi surface pocket), and the antibonding band (which gives rise to the $\beta$ sheet). This inter-band scattering between $\beta$ and $\gamma$ sheets in the pressurized bulk system leads to the $s^\pm$ pairing state that changes sign between these two sheets \cite{Zhang:nc24}. In the present 1UC thin-film case, the stronger $q_z=0$ out-of-plane ferromagnetic fluctuations are intra-band in nature, scattering within the $\gamma$ and within the $\beta$ sheets. This scattering favors a state that changes sign within the $\gamma$ pocket, leading to the $d$-wave states seen in Fig.~\ref{1uc}.

Moreover, for both fillings, $\chi'({\bf q}, \omega=0)$ is strongest for an in-plane ${\bf q}$ close to $ \sim (\pi/3, \pi/3)$, corresponding to a diagonal spin stripe pattern as displayed in Fig.~\ref{Momentum-structure} (c). This unique spin state can be understood in terms of the competition between intraorbital and interorbital hopping mechanisms, as discussed in previous studies~\cite{Lin:prl21,Lin:cp,Lin:prb22,Lin:prb23}. An analysis of the leading contributions to this scattering (see Fig.S9) reveals that it arises primarily from intra-band scattering within the $\gamma$ and $\beta$ pockets, between regions where the Fermi surface Bloch states originate mostly from the Ni2 layer (see Fig.~\ref{Momentum-structure} (d)). This can also be seen in the orbital- and layer-dependent contributions to $\chi'({\bf q},\omega)$ shown in Figs.~\ref{Momentum-structure} (a-b). Here, the dominant contribution arises from the intra-orbital $d_{3z^2-r^2}$ intra-layer scattering $\chi_{33}$ from the Ni2 layer, while the other components are much smaller. The dominant intra-band scattering gives rise to the $d$-wave pairing states shown in Fig.~\ref{1uc}, which change sign within the Fermi surface pockets, rather than between different pockets as in the $s^\pm$ state of the pressurized bulk.

\begin{figure}
\centering
\includegraphics[width=0.44\textwidth]{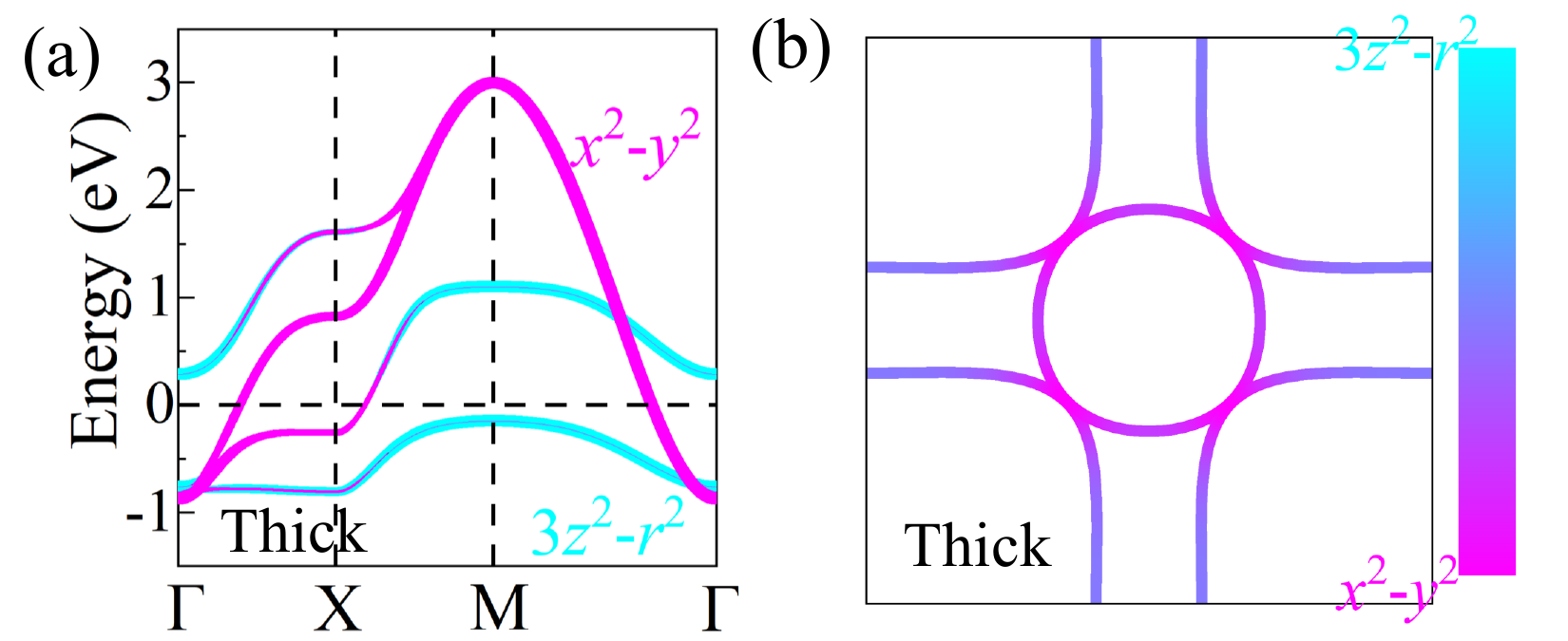}
\caption{ (a) Tight-binding band structures of the thick film grown on an LSAO substrate, and (b) its corresponding Fermi surface at $n = 3.0$ (1.5 per site), respectively.  The hopping files for this thick film are in separate files from the SM~\cite{Supplemental}.}
\label{Thickness}
\end{figure}

{\it Thicker film--} Finally, we briefly discuss a thicker film, dropping the slab geometry. This can be regarded as a bulk system grown on a substrate, by which, the in-plane $(a,b)$ lattice constants are fixed to those of LSAO, but $c$ is allowed to relax. Then, the Ni sites in the BL are symmetric. As shown in Fig.~\ref{Thickness}(a), the $\gamma$ pocket consisting of $d_{3z^2-r^2}$ orbitals shifts further below the Fermi level compared to the 1UC thin film, and this band is flatter than in the 1UC thin film. It also leads to the absence of the $\gamma$ pocket at the Fermi level, as illustrated in Fig.~\ref{Thickness}(b). Thus, to induce the $\gamma$ pocket around the M point, this system needs to be doped with more holes than in the 1UC case, although with the potential doping effect from the substrate likely to be restricted only to the layers close to the substrate and will not affect the overall filling in the bulk. Hence, it is reasonable to expect that thick films would not exhibit superconductivity without pressure, unless additional hole doping is introduced, for example, via La$^{3+}$ to La$^{2+}$ substitution.

{\it Conclusion and discussion--} In summary, we have unveiled strong resemblances but also key differences between strained 1UC thin-film and high-pressure BL nickelates. The ratio of the inter-layer hopping of the $d_{3z^2-r^2}$ orbitals and the intra-layer hopping of the $d_{x^2-y^2}$ orbitals is considerably reduced in the 1UC thin film, compared with the bulk BL nickelate under pressure. Moreover, the crystal-field splitting of the $e_g$ orbitals is also enhanced in the 1UC thin films. These differences cause the downward shift of the $\gamma$ hole pocket around the M point below the Fermi level for the stoichiometric case (electron density $n = 3.0$), leading to negligibly small singlet pairing strength within RPA. Upon hole doping, however, the $\gamma$ pocket appears on the Fermi surface. We find a leading $d_{x^2-y^2}$-wave pairing state at moderate hole doping, followed by a $d_{xy}$-wave pairing state at higher hole doping. The gap is largest on the small hole $\gamma$ pocket at the M point, and changes sign within that pocket. In contrast to the bulk BL nickelates under high-pressure, the superconductivity is characterized by intralayer pairs rather than interlayer pairs, driven by spin-fluctuation scattering primarily from the Ni2 layer, which is the farthest from the substrate. By comparing the size of the $\gamma$ pocket with that observed in recent ARPES measurements for the 1UC sample~\cite{Li:nsr25}, we find that a theoretical model with the electron density per Ni site $n = 2.8$ provides a reasonable agreement, suggesting that the superconducting nickelate BL thin film is indeed {\it hole doped}. Furthermore, in thick films, the crystal-field splitting is similarly increased compared to the high-pressure bulk, but the $\gamma$ pocket is still absent, suggesting that superconductivity may not emerge in this system without additional hole doping. Our results strongly suggest that hole doping is essential to realize superconductivity at ambient pressure in ultra-thin-film BL nickelates.

{\it Acknowledgment--} This work was supported by the U.S. Department of Energy, Office of Science, Basic Energy Sciences, Materials Sciences and Engineering Division.  This manuscript has been authored by UT-Battelle, LLC, under contract DE-AC05-00OR22725 with the US Department of Energy (DOE). The US government retains, and the publisher, by accepting the article for publication, acknowledges that the US government retains a nonexclusive, paid-up, irrevocable, worldwide license to publish or reproduce the published form of this manuscript, or allow others to do so, for US government
purposes. DOE will provide public access to these results
of federally sponsored research in accordance with the DOE
Public Access Plan (https://www.energy.gov/DOE Public Access Plan).

{\it Data availability--} The dataset of the main findings of this study is openly available in a Zenodo Repository. In addition, the hopping and crystal-field parameters for our tight-binding and RPA calculations are available in a separate file of the Supplementary Material and Zenodo Repository for reproducing our results. Simulation RPA codes are available
at https://github.com/maierta/MRPAPP.


\begin{references}
\bibitem{Sun:arxiv} H. Sun, M. Huo, X. Hu, J. Li, Y. Han, L. Tang, Z. Mao, P. Yang, B. Wang, J. Cheng, D.-X. Yao, G.-M. Zhang, and M. Wang, Signatures of superconductivity near 80 K in a nickelate under high pressure \href{https://doi.org/10.1038/s41586-023-06408-7}{Nature \textbf{621} 493 (2023).}
\bibitem{Zhu:arxiv11} Y. Zhu, E. Zhang, B. Pan, X. Chen, D. Peng, L. Chen, H. Ren, F. Liu, N. Li, Z. Xing, J. Han, J. Wang, D. Jia, H. Wo, Y. Gu, Y. Gu, L. Ji, W. Wang, H. Gou, Y. Shen, T. Ying, X. Chen, W. Yang, C. Zheng, Q. Zeng, J.-G. Guo, and J. Zhao, Superconductivity in pressurized trilayer La$_4$Ni$_3$O$_{10-\delta}$ single crystals \href{https://doi.org/10.1038/s41586-024-07553-3}{Nature \textbf{631} 531 (2024).}
\bibitem{Li:cpl} Q. Li, Y.-J. Zhang, Z.-N. Xiang, Y. Zhang, X. Zhu and H.-H. Wen, Signature of Superconductivity in Pressurized La$_4$Ni$_3$O$_{10}$ \href{https://doi.org/10.1088/0256-307X/41/1/017401}{Chinese Phys. Lett. \textbf{41}, 017401 (2024).}
\bibitem{Zhang:arxiv-exp} Y. Zhang, D. Su, Y. Huang, H. Sun, M. Huo, Z. Shan, K. Ye, Z. Yang, R. Li, M. Smidman, M. Wang, L. Jiao, and H. Yuan, High-temperature superconductivity with zero resistance and strange-metal behaviour in La$_3$Ni$_2$O$_{7-\delta}$ \href{https://doi.org/10.1038/s41567-024-02515-y}{Nat. Phys. \textbf{20} 1269 (2024).}
\bibitem{Hou:arxiv} J. Hou, P. T. Yang, Z. Y. Liu, J. Y. Li, P. F. Shan, L. Ma, G. Wang, N. N. Wang, H. Z. Guo, J. P. Sun, Y. Uwatoko, M. Wang, G.-M. Zhang, B. S. Wang, and J.-G. Cheng, Emergence of High-Temperature Superconducting Phase in Pressurized La$_3$Ni$_2$O$_7$ Crystals \href{https://doi.org/10.1088/0256-307X/40/11/117302}{Chin. Phys. Lett. \textbf{40} 117302 (2023).}
\bibitem{Sakakibara:arxiv09} H. Sakakibara, M. Ochi, H. Nagata, Y. Ueki, H. Sakurai, R. Matsumoto, K. Terashima, K. Hirose, H. Ohta, M. Kato, Y. Takano, and K. Kuroki, Theoretical analysis on the possibility of superconductivity in the trilayer Ruddlesden-Popper nickelate La$_4$Ni$_3$O$_{10}$ under pressure and its experimental examination: Comparison with La$_3$Ni$_2$O$_7$ \href{https://doi.org/10.1103/PhysRevB.109.144511}{Phys. Rev. B \textbf{109} 144511 (2024).}
\bibitem{Wang:arxiv9} G. Wang, N. N. Wang, X. L. Shen, J. Hou, L. Ma, L. F. Shi, Z. A. Ren, Y. D. Gu, H. M. Ma, P. T. Yang, Z. Y. Liu, H. Z. Guo, J. P. Sun, G. M. Zhang, S. Calder, J.-Q. Yan, B. S. Wang, Y. Uwatoko, and J.-G. Cheng, Pressure-Induced Superconductivity In Polycrystalline La$_3$Ni$_2$O$_{7-\delta}$ \href{https://doi.org/10.1103/PhysRevX.14.011040}{Phys. Rev. X \textbf{14} 011040 (2024).}
\bibitem{Wang:nature} N. Wang, G. Wang, X. Shen, J. Hou, J. Luo, X. Ma, H. Yang, L. Shi, J. Dou, J. Feng, J. Yang, Y. Shi, Z. Ren, H. Ma, P. Yang, Z. Liu, Y. Liu, H. Zhang, X. Dong, Y. Wang, K. Jiang, J. Hu, S. Nagasaki, K. Kitagawa, S. Calder, J. Yan, J. Sun, B. Wang, R. Zhou, Y. Uwatoko, and J. Cheng, Bulk high-temperature superconductivity in pressurized tetragonal La$_2$PrNi$_2$O$_7$ \href{https://doi.org/10.1038/s41586-024-07996-8}{Nature \textbf{634}, 579 (2024).}
\bibitem{Zhang:arxiv11} M. Zhang, C. Pei, X. Du, Y. Cao, Q. Wang, J. Wu, Y. Li, Y. Zhao, C. Li, W. Cao, S. Zhu, Q. Zhang, N. Yu, P. Cheng, J. Zhao, Y. Chen, H. Guo, L. Yang, and Y. Qi, Superconductivity in trilayer nickelate La$_4$Ni$_3$O$_{10}$ under pressure \href{https://doi.org/10.1103/PhysRevX.15.021005}{Phys. Rev. X \textbf{15}, 021005 (2025).}
\bibitem{Zhang:prb23} Y. Zhang, L.-F. Lin, A. Moreo, and E. Dagotto, Electronic structure, dimer physics, orbital-selective behavior, and magnetic tendencies in the bilayer nickelate superconductor La$_3$Ni$_2$O$_7$ under pressure \href{https://doi.org/10.1103/PhysRevB.108.L180510}{Phys. Rev. B \textbf{108}, L180510 (2023).}
\bibitem{LiuZhe:arxiv} Z. Liu, M. Huo, J. Li, Q. Li, Y. Liu, Y. Dai, X. Zhou, J. Hao, Y. Lu, M. Wang, and W.-H. Wen, Electronic correlations and partial gap in the bilayer nickelate La$_3$Ni$_2$O$_7$ \href{https://doi.org/10.1038/s41467-024-52001-5}{Nat. Commun. \textbf{15} 7570 (2024).}
\bibitem{Yang:arxiv09} J. Yang, H. Sun, X. Hu, Y. Xie, T. Miao, H. Luo, H. Chen, B. Liang, W. Zhu, G. Qu, C.-Q. Chen, M. Huo, Y. Huang, S. Zhang, F. Zhang, F. Yang, Z. Wang, Q. Peng, H. Mao, G. Liu, Z. Xu, T. Qian, D.-X. Yao, M. Wang, L. Zhao, and X. J. Zhou, Orbital-dependent electron correlation in double-layer nickelate La$_3$Ni$_2$O$_7$ \href{https://doi.org/10.1038/s41467-024-48701-7}{Nat. Commun. \textbf{15} 4373 (2024).}
\bibitem{Zhang:prb24} Y. Zhang, L.-F. Lin, A. Moreo, T. A. Maier, and E. Dagotto, Electronic structure, magnetic correlations, and superconducting pairing in the reduced Ruddlesden-Popper bilayer La$_3$Ni$_2$O$_6$  under pressure: Different role of $d_{3z^2-r^2}$ orbital compared with La$_3$Ni$_2$O$_7$ \href{https://doi.org/10.1103/PhysRevB.109.045151}{Phys. Rev. B \textbf{109}, 045151 (2024).}
\bibitem{Dong:arxiv12} Z. Dong, M. Huo, J. Li, J. Li, P. Li, H. Sun, Y. Lu, M. Wang, Y. Wang, and Z. Chen, Visualization of oxygen vacancies and self-doped ligand holes in La$_3$Ni$_2$O$_{7-\delta}$ \href{https://doi.org/10.1038/s41586-024-07482-1}{Nature \textbf{630} 847 (2024).}
\bibitem{Xie:SB} T. Xie, M. Huo, X. Ni, F. Shen, X. Huang, H. Sun, H. C. Walker, D. Adroja, D. Yu, B. Shen, L. He, K. Cao, and M. Wang, Strong interlayer magnetic exchange coupling in La$_3$Ni$_2$O$_{7-\delta}$ revealed by inelastic neutron scattering \href{https://doi.org/10.1016/j.scib.2024.07.030}{Science Bulletin \textbf{69} 3221 (2024).}
\bibitem{Chen:arxiv2024} X. Chen, J. Choi, Z. Jiang, J. Mei, K. Jiang, J. Li, S. Agrestini, M. Garcia-Fernandez, X. Huang, H. Sun, D. Shen, M. Wang, J. Hu, Y. Lu, K.-J. Zhou, and D. Feng, Electronic and magnetic excitations in La$_3$Ni$_2$O$_7$ \href{https://doi.org/10.1038/s41467-024-53863-5}{Nat. Commun. \textbf{15}, 9597 (2024).}
\bibitem{Dan:arxiv2024} Z. Dan, Y. Zhou, M. Huo, Y. Wang, L. Nie, M. Wang, T. Wu, and X. Chen,  Pressure-enhanced spin-density-wave transition in double-layer nickelate La$_3$Ni$_2$O$_{7-\delta}$ \href{https://doi.org/10.1016/j.scib.2025.02.019}{Science Bulletin \textbf{70}, 1239 (2025).}
\bibitem{Zhang:prb23-2} Y. Zhang, L.-F. Lin, A. Moreo, T. A. Maier, and E. Dagotto, Trends in electronic structures and ${s^\pm}$-wave pairing for the rare-earth series in bilayer nickelate superconductors $R_3$Ni$_2$O$_7$ \href{https://doi.org/10.1103/PhysRevB.108.165141}{Phys. Rev. B \textbf{108}, 165141 (2023).}
\bibitem{Chen:prl24} K. Chen, X. Liu, J. Jiao, M. Zou, C. Jiang, X. Li, Y. Luo, Q. Wu, N. Zhang, Y. Guo, and L. Shu, Evidence of Spin Density Waves in La$_3$Ni$_2$O$_{7-\delta}$ \href{https://doi.org/10.1103/PhysRevLett.132.256503}{Phys. Rev. Lett. \textbf{132}, 256503 (2024).}
\bibitem{Sakakibara:prl24} H. Sakakibara, N. Kitamine, M. Ochi, and K. Kuroki, Possible High $T_c$ Superconductivity in La$_3$Ni$_2$O$_7$ under High Pressure through Manifestation of a Nearly Half-Filled BL Hubbard Model \href{https://doi.org/10.1103/PhysRevLett.132.106002}{Phys. Rev. Lett. \textbf{132} 106002 (2024).}
\bibitem{Lin:prb23} L.-F. Lin, Y. Zhang, N. Kaushal, G. Alvarez, T. A. Maier, A. Moreo, and E. Dagotto, Magnetic phase diagram of a two-orbital model for bilayer nickelates with varying doping \href{https://doi.org/10.1103/PhysRevB.110.195135}{Phys. Rev. B \textbf {110}, 195135 (2024).}
\bibitem{Oh:prb24} H. Oh, B. Zhou, and Y.-H. Zhang, Type-II $t-J$ model in charge transfer regime in bilayer La$_3$Ni$_2$O$_7$ and trilayer La$_4$Ni$_3$O$_{10}$ \href{https://doi.org/10.1103/PhysRevB.110.195135}{Phys. Rev. B \textbf {110}, 195135 (2024).}
\bibitem{Li:Nature} D. Li, K. Lee, B. Y. Wang, M. Osada, S. Crossley, H. R. Lee, Y. Cui, Yi, Y. Hikita,and H. Y. Hwang, Superconductivity in an infinite-layer nickelate \href{https://doi.org/10.1038/s41586-019-1496-5}{Nature \textbf{572}, 624 (2019).}
\bibitem{Zhang:prb20} Y. Zhang, L.-F. Lin, W. Hu, A. Moreo, S. Dong, and E. Dagotto, Similarities and differences between nickelate and cuprate films grown on a SrTiO$_3$ substrate \href{https://doi.org/10.1103/PhysRevB.102.195117}{Phys. Rev. B \textbf{102}, 195117 (2020).}
\bibitem{Nomura:rpp} Y. Nomura and R. Arita, Superconductivity in infinite-layer nickelates\href{https://doi.org/10.1088/1361-6633/ac5a60}{ Rep. Prog. Phys. \textbf{85}, 052501  (2022).}
\bibitem{Pan:nm} G. A. Pan, D. F. Segedin, H. LaBollita, Q. Song, E. M. Nica, B. H. Goodge, A. T. Pierce, S. Doyle, S. Novakov, D. C. Carrizales, A. T. N'Diaye, P. Shafer, H. Paik, J. T. Heron, J. A. Mason, A. Yacoby, L. F. Kourkoutis, O. Erten, C. M. Brooks, A. S. Botana and J. A. Mundy, Superconductivity in a quintuple-layer square-planar nickelate \href{https://doi.org/10.1038/s41563-021-01142-9}{Nat. Mater.  \textbf{21}, 160 (2022).}
\bibitem{Chen:prb24} J. Chen, F. Yang, and W. Li, Spin-density wave and superconductivity in La$_4$Ni$_3$O$_{10}$ under ambient pressure \href{https://doi.org/10.1103/PhysRevB.110.L041111}{Phys. Rev. B \textbf {110}, L041111 (2024).}
\bibitem{Ryee:prl} S. Ryee, N. Witt, and T. O. Wehling, Quenched Pair Breaking by Interlayer Correlations as a Key to Superconductivity in La$_3$Ni$_2$O$_7$ \href{https://doi.org/10.1103/PhysRevLett.133.096002}{Phys. Rev. Lett. \textbf {133}, 096002 (2024).}
\bibitem{Zhang:prl} J. Zhan, Y. Gu, X. Wu, and J. Hu, Cooperation between Electron-Phonon Coupling and Electronic Interaction in bilayer Nickelates La$_3$Ni$_2$O$_7$ \href{https://doi.org/10.1103/PhysRevLett.134.136002}{Phys. Rev. Lett. \textbf {134}, 136002 (2025).}
\bibitem{LaBollita:prm} H. LaBollita, V. Pardo, M. R. Norman, and A. S. Botana, Assessing spin-density wave formation in La$_3$Ni$_2$O$_7$ from electronic structure calculations
 \href{https://doi.org/10.1103/PhysRevMaterials.8.L111801}{Phys. Rev. Mater. \textbf {8}, L111801 (2025).}
\bibitem{Lu:prb25} M. Lu, and T. Zhou, Assessing spin-density wave formation in La$_3$Ni$_2$O$_7$ from electronic structure calculations
 \href{https://doi.org/10.1103/PhysRevB.111.094504}{Phys. Rev. B \textbf {111}, 094504 (2025).}
\bibitem{Duan:arxiv25} G. Duan, Z. Liao, L. Chen, Y. Wang, R. Yu, and Q. Si, Orbital-selective correlation effects and superconducting pairing symmetry in a multiorbital $t-J$ model for bilayer nickelates\href{https://doi.org/10.48550/arXiv.2502.09195}{arXiv 2502.09195 (2025).}
\bibitem{Braz:arxiv25} L. B. Braz, G. B. Martins, and L. G. G. V. Dias da Silva, Interlayer interactions in La$_3$Ni$_2$O$_7$ under pressure: from $s^\pm$ to $d_{xy}$-wave superconductivity\href{https://doi.org/10.1103/f4wf-56fl}{Phys. Rev. Research \textbf {7}, 033023 (2025).}
\bibitem{Luo:prl23} Z. Luo, X. Hu, M. Wang, W. Wu, and D.-X. Yao, Bilayer Two-Orbital Model of La$_3$Ni$_2$O$_7$ under Pressure \href{https://doi.org/10.1103/PhysRevLett.131.126001}{Phys. Rev. Lett. \textbf{131}, 126001 (2023).}
\bibitem{Zhang:arxiv24} Y. Zhang, L.-F. Lin, A. Moreo, T. A. Maier, and E. Dagotto, Prediction of $s^{\pm}$-wave superconductivity enhanced by electronic doping in trilayer nickelates La$_4$Ni$_3$O$_{10}$ under pressure \href{https://doi.org/10.1103/PhysRevLett.133.136001}{Phys. Rev. Lett. \textbf{133}, 136001 (2024).}
\bibitem{LaBollita:prb24} H. LaBollita, J. Kapeghian, M. R. Norman, and A. S. Botana, Electronic structure and magnetic tendencies of trilayer La$_4$Ni$_3$O$_{10}$ under pressure: Structural transition, molecular orbitals, and layer differentiation \href{https://doi.org/10.1103/PhysRevB.109.195151}{Phys. Rev. B \textbf{109}, 195151 (2024).}
\bibitem{Qu:prl} X.-Z. Qu, D.-W. Qu, J. Chen, C. Wu, F. Yang, W. Li, and G. Su, BL $t-J-J_{\perp}$ Model and Magnetically Mediated Pairing in the Pressurized Nickelate La$_3$Ni$_2$O$_7$ \href{https://doi.org/10.1103/PhysRevLett.132.036502}{Phys. Rev. Lett. \textbf{132}, 036502 (2024).}
\bibitem{Lu:prl} C. Lu, Z. Pan, F. Yang, and C. Wu, Interlayer-Coupling-Driven High-Temperature Superconductivity in La$_3$Ni$_2$O$_7$ under Pressure \href{https://doi.org/10.1103/PhysRevLett.132.146002}{Phys. Rev. Lett. \textbf{132}, 146002 (2024).}
\bibitem{Tian:prb24} Y.-H. Tian, Y. Chen, J.-M. Wang, R.-Q. He, and Z.-Y. Lu, Correlation effects and concomitant two-orbital $s_{\pm}$-wave superconductivity in La$_3$Ni$_2$O$_7$ under high pressure \href{https://doi.org/10.1103/PhysRevB.109.165154}{Phys. Rev. B \textbf{109}, 165154 (2024).}
\bibitem{Liu:arxiv2023} H. Liu, C. Xia, S. Zhou, H. Chen, Sensitive dependence of pairing symmetry on Ni-eg crystal field splitting in the nickelate superconductor La$_3$Ni$_2$O$_7$ \href{https://doi.org/10.1038/s41467-025-56206-0}{Nat. Commun. \textbf{16}, 1054 (2025).}
\bibitem{Zhang:2323} Y. Zhang, L.-F. Lin, A. Moreo, T. A. Maier, and E. Dagotto, Magnetic Correlations and Pairing Tendencies of the Hybrid Stacking Nickelate Superlattice La$_7$Ni$_5$O$_{17}$ (La$_3$Ni$_2$O$_7$/La$_3$Ni$_4$O$_{10}$)under Pressure \href{https://doi.org/10.1103/t4zz-zbw1}{Phys. Rev. B \textbf{112}, 024508 (2025).}
\bibitem{Shi:arxiv25} M. Shi, D. Peng, K. Fan, Z. Xing, S. Yang, Y. Wang, H. Li, R. Wu, M. Du, B. Ge, Z. Zeng, Q. Zeng, J. Ying, T. Wu, X. Chen, Superconductivity of the hybrid Ruddlesden-Popper La$_5$Ni$_3$O$_{11}$ single crystals \href{https://doi.org/10.1038/s41567-025-03023-3}{Nat. Phys. \textbf{24}, 1780 (2025).}
\bibitem{Zhang:1212} Y. Zhang, L.-F. Lin, A. Moreo, S. Okamoto, T. A. Maier, and E. Dagotto, Electronic Structure, Magnetic and Pairing Tendencies of Alternating Single-layer BL Stacking Nickelate La$_5$Ni$_3$O$_{11}$ Under Pressure \href{https://doi.org/10.1103/1mr2-s6s8}{Phys. Rev. B \textbf{112}, 094515 (2025).}
\bibitem{Zhang:1313} Y. Zhang, L.-F. Lin, A. Moreo, T. A. Maier, and E. Dagotto, Electronic structure, self-doping, and superconducting instability in the alternating single-layer trilayer stacking nickelates La$_3$Ni$_2$O$_7$ \href{https://doi.org/10.1103/PhysRevB.110.L060510}{Phys. Rev. B \textbf{110} L060510  (2024).}
\bibitem{Takegami:prb} D. Takegami, K. Fujinuma, R. Nakamura, M. Yoshimura, K.-D. Tsuei, G. Wang, N. N. Wang,
J.-G. Cheng, Y. Uwatoko, and T. Mizokawa, Absence of Ni$^{2+}$/Ni$^{3+}$ charge disproportionation and possible roles of O $2p$ holes in La$_3$Ni$_2$O$_{7-\delta}$ revealed by hard x-ray photoemission spectroscopy \href{https://doi.org/10.1103/PhysRevB.109.125119}{Phys. Rev. B \textbf{109}, 125119 (2024).}
\bibitem{Xu:prb25} S. Xu, C.-Q. Chen, M. Huo, D. Hu, H. Wang, Q. Wu, R. Li, D. Wu, M. Wang, D.-X. Yao, T. Dong, and N. Wang, Origin of the density wave instability in trilayer nickelate La$_4$Ni$_3$O$_{10}$ revealed by optical and ultrafast spectroscopy \href{https://doi.org/10.1103/PhysRevB.111.075140}{Phys. Rev. B \textbf{111}, 075140 (2025).}
\bibitem{Zhang:nc24} Y. Zhang, L.-F. Lin, A. Moreo, T. A. Maier, and E. Dagotto, Structural phase transition, $s_{\pm}$-wave pairing, and magnetic stripe order in BLed superconductor La$_3$Ni$_2$O$_7$ under pressure \href{https://doi.org/10.1038/s41467-024-46622-z}{Nat. Commun. \textbf{15}, 2470 (2024).}
\bibitem{Christiansson:arxiv} V. Christiansson, F. Petocchi and P. Werner, Correlated Electronic Structure of La$_3$Ni$_2$O$_7$ under Pressure \href{https://doi.org/10.1103/PhysRevLett.131.206501}{Phys. Rev. Lett. \textbf{131},  206501 (2023).}
\bibitem{Yang:arxiv} Q.-G. Yang, D. Wang, and Q.-H. Wang, Possible ${s^\pm}$-wave superconductivity in La$_3$Ni$_2$O$_7$ \href{https://doi.org/10.1103/PhysRevB.108.L140505}{Phys. Rev. B \textbf{108}, L140505 (2023).}
\bibitem{Liu:arxiv} Y.-B. Liu, J.-W. Mei, F. Ye, W.-Q. Chen, and F. Yang, ${s^\pm}$-Wave Pairing and the Destructive Role of Apical-Oxygen Deficiencies in La$_3$Ni$_2$O$_7$ under Pressure \href{https://doi.org/10.1103/PhysRevLett.131.236002}{Phys. Rev. Lett. \textbf{131}, 236002 (2023).}
\bibitem{Yang:arxiv1} Y.-F. Yang, . G.-M. Zhang, and F.-C. Zhang, Interlayer valence bonds and two-component theory for high-$T_c$ superconductivity of La$_3$Ni$_2$O$_7$ under pressure \href{https://doi.org/10.1103/PhysRevB.108.L201108}{Phys. Rev. B \textbf{108}, L201108 (2023).}
\bibitem{Oh:arxiv} H. Oh and Y. H. Zhang, Type-II $t-J$ model and shared superexchange coupling from Hund's rule in superconducting La$_3$Ni$_2$O$_7$ \href{https://doi.org/10.1103/PhysRevB.108.174511}{Phys. Rev. B \textbf{108}, 174511 (2023).}
\bibitem{Liao:arxiv} Z. Liao, L. Chen, G. Duan, Y. Wang, C. Liu, R. Yu, and Q. Si, Electron correlations and superconductivity in La$_3$Ni$_2$O$_7$ under pressure tuning \href{https://doi.org/10.1103/PhysRevB.108.214522}{Phys. Rev. B \textbf{108}, 214522 (2023).}
\bibitem{Fan:STM} S. Fan, M. Ou, M. Scholten, Q. Li, Z. Shang, Y. Wang, J. Xu, H. Yang, I. M. Eremin, H.-H. Wen, Superconducting gaps revealed by STM measurements on
La$_2$PrNi$_2$O$_7$ thin films at ambient pressure \href{https://arxiv.org/abs/2506.01788}{arXiv 2506.01788 (2025).}
\bibitem{Lechermann:arxiv} F. Lechermann, J. Gondolf, S. B\"otzel, and I. M. Eremin, Electronic correlations and superconducting instability in La$_3$Ni$_2$O$_7$ under high pressure \href{https://doi.org/10.1103/PhysRevB.108.L201121}{Phys. Rev. B \textbf{108}, L201121 (2023).}
\bibitem{Shilenko:arxiv} D. A. Shilenko, and I. V. Leonov, Correlated electronic structure, orbital-selective behavior, and magnetic correlations in double-layer under pressure \href{https://doi.org/10.1103/PhysRevB.108.125105}{Phys. Rev. B \textbf{108}, 125105 (2023).}
\bibitem{Huang:arxiv} J. Huang, Z. D. Wang, and T. Zhou, Impurity and vortex states in the BL high-temperature superconductor La$_3$Ni$_2$O$_7$ \href{https://doi.org/10.1103/PhysRevB.108.174501}{Phys. Rev. B \textbf{108}, 174501 (2023).}
\bibitem{Qin:arxiv} Q. Qin, and Y.-F. Yang, High-$T_c$ superconductivity by mobilizing local spin singlets and possible route to higher $T_c$ in pressurized La$_3$Ni$_2$O$_7$ \href{https://doi.org/10.1103/PhysRevB.108.L140504}{Phys. Rev. B \textbf{108}, L140504 (2023).}
\bibitem{Wang:PRB24} Q.-G. Yang, K.-Y. Jiang, D. Wang, H.-Y. Lu, and Q.-H. Wang, Effective model and ${s^\pm}$-wave superconductivity in trilayer nickelate La$_4$Ni$_3$O$_{10}$ \href{https://doi.org/10.1103/PhysRevB.109.165140}{Phys. Rev. B \textbf{109}, 165140 (2024).}
\bibitem{Cao:arxiv} Y. Cao, and Y.-F. Yang,  Flat bands promoted by Hund's rule coupling in the candidate double-layer high-temperature superconductor La$_3$Ni$_2$O$_7$ \href{https://doi.org/10.1103/PhysRevB.109.L081105}{Phys. Rev. B \textbf{109}, L081105 (2024).}
\bibitem{Maier:arxiv25} T. A. Maier, P. Doak, L.-F. Lin, Y. Zhang, A. Moreo, and E. Dagotto, Interlayer Pairing in bilayer Nickelates \href{https://doi.org/10.48550/arXiv.2506.07741}{arXiv:2506.07741 (2025).}
\bibitem{Li:scpma} J. Li, C. Chen, C. Huang, Y. Han, M. Huo, X. Huang, P. Ma, Z. Qiu, J. Chen, X. Hu, L. Chen, T. Xie, B. Shen, H. Sun, D. Yao, and M. Wang, \href{https://doi.org/10.1007/s11433-023-2329-x}{Sci. China Phys. Mech. Astron. \textbf{67}, 117403 (2024).}
\bibitem{Li:arxiv25} F. Li, D. Peng, J. Dou, N. Guo, L. Ma, C. Liu, L. Wang, Y. Zhang, J. Luo, J. Yang, J. Zhang, W. Cai, J. Cheng, Q. Zheng, R. Zhou, Q. Zeng, X. Tao, and J. Zhang, Ambient pressure growth of bilayer nickelate single crystals with superconductivity over 90 K under high pressure \href{
https://doi.org/10.48550/arXiv.2501.14584}{arXiv:2501.14584 (2025).}
\bibitem{Lip:nsr25} J.  Li, P. Ma, H. Zhang, X. Huang, C. Huang, M. Huo, D.Hu, Z. Dong, C. He, J. Liao, X. Chen, T. Xie, H. Sun, M. Wang,  Identification of superconductivity in bilayer nickelate La$_3$Ni$_2$O$_7$ under high pressure up to 100 GPa \href{https://doi.org/10.1093/nsr/nwaf220}{ Nat. Sci. Rev., nwaf220 (2025).}
\bibitem{QiuZ:arxiv25} Z. Qiu, J. Chen, D. V. Semenok, Q. Zhong, D. Zhou, J. Li, P. Ma, X. Huang, M. Huo, T. Xie, X. Chen, H.-k. Mao, V. Struzhkin, H. Sun, and M. Wang, Interlayer coupling enhanced superconductivity near 100 K in La$_{3-x}$Nd$_x$Ni$_2$O$_7$ \href{https://doi.org/10.48550/arXiv.2510.12359}{arXiv:2510.12359 (2025).}
\bibitem{Guo:arxiv25} J. Guo, Y. Chen, Y. Wang, H. Sun, D. Hu, M. Wang, X. Huang, and T. Cu, Revealing superconducting gap in La$_3$Ni$_2$O$_{7-\delta}$ by Andreev reflection spectroscopy under high pressure \href{https://doi.org/10.48550/arXiv.2509.12601}{arXiv:2509.12601 (2025).}
\bibitem{Dong:arxiv25} Z. Dong, G. Wang, N. Wang, W.-H. Dong, L. Gu, Y. Xu, J. Cheng, Z. Chen, Y. Wang, Interstitial oxygen order and its competition with superconductivity in La$_2$PrNi$_2$O$_{7+\delta}$ \href{https://doi.org/10.48550/arXiv.2508.03414}{arXiv:2508.03414 (2025).}
\bibitem{Ko:nature} E. K. Ko, Y. Yu, Y. Liu, L. Bhatt, J. Li, V. Thampy, C.-T. Kuo, B. Y. Wang, Y. Lee, K. Lee, J.-S. Lee, B. H. Goodge, D. A. Muller, and H. Y. Hwang, Signatures of ambient pressure superconductivity in thin film La$_3$Ni$_2$O$_7$ \href{https://doi.org/10.1038/s41586-024-08525-3}{Nature \textbf{638}, 935 (2025).}
\bibitem{Liu:nm25} Y. Liu, E. K. Ko, Y. Tarn, L. Bhatt, J. Li, V. Thampy,  B. H. Goodge, D. A. Muller, S. Raghu, Y. Yu, and H. Y. Hwang, Superconductivity and normal-state transport in compressively strained La$_2$PrNi$_2$O$_7$ thin films \href{https://doi.org/10.1038/s41563-025-02258-y}{Nat, Mater. \textbf{24}, 1221 (2025).}
\bibitem{Zhou:nature} G. Zhou, W. Lv, H. Wang, Z. Nie, Y. Chen, Y. Li, H. Huang, W.-Q. Chen, Y.-J. Sun, Q.-K. Xue and Z. Chen Ambient-pressure superconductivity onset above 40 K in (La,Pr)$_3$Ni$_2$O$_7$ films \href{https://doi.org/10.1038/s41586-025-08755-z}{Nature \textbf{640}, 641 (2025).}
\bibitem{Geisler:arxiv25} B. Geisler, J. J. Hamlin, G. R. Stewart, R. G. Hennig, and P.J. Hirschfeld, Electronic reconstruction and interface engineering of emergent spin fluctuations in compressively
strained La$_3$Ni$_2$O$_7$ on SrLaAlO$_4$(001) \href{https://doi.org/10.48550/arXiv.2503.10902}{arXiv:2502.17831 (2025).}
\bibitem{Huang:arxiv25} J. Huang and T. Zhou, $s^{\pm}$ pairing via interlayer interaction in La$_{2.85}$Pr$_{0.15}$Ni$_2$O$_7$ Thin Films under Ambient Pressure \href{https://doi.org/10.48550/arXiv.2503.10902}{arXiv:2502.17831 (2025).}
\bibitem{Hao:arxiv25} B. Hao, M. Wang, W. Sun, Y. Yang, Z. Mao, S. Yan, H. Sun, H. Zhang, L. Han, Z. Gu, J. Zhou, D. Ji, and Y. Nie, Superconductivity and phase diagram in Sr-doped La$_3$Ni$_2$O$_7$ thin films
 \href{https://doi.org/10.48550/arXiv.2505.12603}{arXiv:2505.12603 (2025).}
\bibitem{Xiong:arxiv25} Y. Xiong, Y. Cai, and T. Ma, Pairing Symmetry Crossover from $d$-wave to $s^{\pm}$-wave in a bilayer Nickelate Driven by
Hund’s Coupling and Crystal Field Splitting \href{https://doi.org/10.48550/arXiv.2510.19406}{arXiv:2510.19406 (2025).}
\bibitem{Ushio:arxiv25} K. Ushio, S. Kamiyama, Y. Hoshi, R. Mizuno, M. Ochi, K. Kuroki, and H. Sakakibara, Theoretical study on ambient pressure superconductivity in La$_3$Ni$_2$O$_7$ thin films :
structural analysis, model construction, and robustness of $s^{\pm}$-wave pairing \href{https://doi.org/10.48550/arXiv.2506.20497}{arXiv:2506.20497 (2025).}
\bibitem{Qiu:arxiv25} W. Qiu, Z. Luo, X. Hu, and D.-X. Yao, Pairing symmetry and superconductivity in La$_3$Ni$_2$O$_7$ thin films\href{https://doi.org/10.48550/arXiv.2506.20727} {arXiv:2506.20727 (2025).}
\bibitem{Cao:arxiv25} Y.-H. Cao, K.-Y. Jiang, H.-Y. Lu, D. Wang, and Q.-H. Wang, Strain-Engineered Electronic Structure and Superconductivity in La$_3$Ni$_2$O$_7$ Thin Films \href{https://doi.org/10.48550/arXiv.2507.13694}{arXiv:2507.13694 (2025).}
\bibitem{Shao:arxiv25} Z.-Y. Shao, C. Lu, M. Liu, Y.-B. Liu, Z. Pan, C. Wu, and F. Yang, Pairing without $\gamma$-Pocket in the La$_3$Ni$_2$O$_7$ Thin Films \href{https://doi.org/10.48550/arXiv.2507.20287}{arXiv:2507.20287 (2025).}
\bibitem{Xiang:arxiv25} L. Xiang, S. Lei, X. Ren, Z. Han, Z. Xu, X. J. Zhou, and Z. Zhu, Stabilizing and Tuning Superconductivity in La$_3$Ni$_2$O$_{7-\delta}$ Films: Oxygen Recycling Protocol Reveals Hole-Doping Analogue \href{https://doi.org/10.48550/arXiv.2508.11581}{arXiv:2508.11581 (2025).}
\bibitem{Ji:arxiv25} H. Ji, Z. Xie, Y. Chen, G. Zhou, L. Pan, H. Wang, H. Huang, J. Ge, Y. Liu, G.-M. Zhang, Z. Wang, Q.-K. Xue, Z. Chen, and J. Wang, Signatures of spin-glass superconductivity in nickelate (La, Pr, Sm)$_3$Ni$_2$O$_7$ films \href{https://doi.org/10.48550/arXiv.2508.16412}{arXiv:2508.16412 (2025).}
\bibitem{Lv:arxiv25} W. Lv, Z. Nie, H. Wang, H. Huang, Q. Xue, G. Zhou, Z. Chen, Growth optimization of Ruddlesden-Popper nickelate high-temperature superconducting thin films \href{https://doi.org/10.48550/arXiv.2508.18107}{arXiv:2508.18107 (2025).}
\bibitem{Bleys:arxiv25} L. Bleys, N. Corkill, Y.-F. Zhao, G. L. Pascut, H. LaBollita, A. S. Botana, and K. F. Quader, Role of correlations in Ruddlesden-Popper bilayer nickelates under compressive strain \href{https://doi.org/10.48550/arXiv.2509.00940}{arXiv:2509.00940 (2025).}
\bibitem{Liu:arxiv25} L. Liu, J. Peng, Z. Qiao, S. Cai, H. Dong, Y. Jia, and Z. Zhang, Optimally Tensile Strained La$_3$Ni$_2$O$_7$ Films as Candidate High-
Temperature Superconductors on Designer Ba$_{1-x}$Sr$_x$O (001) and SrOSrTiO$_3$ Substrates \href{https://doi.org/10.48550/arXiv.2509.13820}{arXiv:2509.13820 (2025).}
\bibitem{Osada:cp25} M. Osada, C. Terakura, A. Kikkawa, M. Nakajima, H.-Y. Chen, Y. Nomura, Y. Tokura, and A. Tsukazaki, Strain-tuning for superconductivity in La$_3$Ni$_2$O$_7$ thin films \href{https://doi.org/10.1038/s42005-025-02154-6}{Commun. Phys. \textbf{8} 251 (2025).}
\bibitem{Tarn:arxiv25} Y. Tarn, Y. Liu, F. Theuss, J. Li, B. Y. Wang, J. Wang, V. Thampy, Z.-X. Shen, Y. Yu, and H. Y. Hwang, Reducing the strain required for ambient-pressure superconductivity in bilayer nickelates \href{https://doi.org/10.48550/arXiv.2510.27613}{arXiv:2510.27613 (2025).}
\bibitem{Shen:arxiv25} J. Shen, G. Zhou, Y. Miao, P. Li, Z. Ou, Y. Chen, Z. Wang, R. Luan, H. Sun, Z. Feng, X. Yong, Y. Li, L. Xu, W. Lv, Z. Nie, H. Wang, H. Huang, Y.-J. Sun, Q.-K. Xue, J. He, and Z. Chen, Nodeless superconducting gap and electron-boson coupling in (La,Pr,Sm)$_3$Ni$_2$O$_7$ \href{https://doi.org/10.48550/arXiv.2502.17831}{arXiv:2502.17831 (2025).}
\bibitem{Wang:arxiv25} B. Y. Wang, Y. Zhong, S. Abadi, Y. Liu, Y. Yu, X. Zhang, Y.-M. Wu, R. Wang, J. Li, Y. Tarn, E. K. Ko, V. Thampy, M. Hashimoto, D. Lu, Y. S. Lee, T. P. Devereaux, C. Jia, H. Y. Hwang, and Z.-X. Shen, Electronic structure of compressively strained thin film La$_2$PrNi$_2$O$_7$ \href{https://doi.org/10.48550/arXiv.2504.16372}{arXiv:2504.16372 (2025).}
\bibitem{Li:nsr25} P. Li, G. Zhou, W. Lv, Y. Li, C. Yue, H. Huang, L. Xu, J. Shen, Y. Miao, W. Song, Z. Nie, Y. Chen, H. Wang, W. Chen, Y. Huang, Z.-H. Chen, T. Qian, J. Lin, J. He, Y.-J. Sun, Z. Chen , Q.-K. Xue, Angle-resolved photoemission spectroscopy of superconducting (La,Pr)$_3$Ni$_2$O$_7$/SrLaAlO$_4$ heterostructures \href{https://doi.org/10.1093/nsr/nwaf205}{Natl. Sci. Rev. nwaf205 (2025).}
\bibitem{Kresse:Prb} G. Kresse and J. Hafner, Ab initio molecular dynamics for liquid metals \href{https://doi.org/10.1103/PhysRevB.47.558}{Phys. Rev. B \textbf{47}, 558 (1993).}
\bibitem{Kresse:Prb96} G.~Kresse and J.~Furthm\"{u}ller, Generalized Gradient Approximation Made Simple \href{https://doi.org/10.1103/PhysRevB.54.11169}{Phys. Rev. B \textbf{54}, 11169 (1996).}
\bibitem{Blochl:Prb} P. E. Bl\"{o}chl, Projector augmented-wave method \href{https://doi.org/10.1103/PhysRevB.50.17953}{Phys. Rev. B \textbf{50}, 17953 (1994).}
\bibitem{Perdew:Prl} J. P. Perdew, K. Burke, and M. Ernzerhof, Generalized Gradient Approximation Made Simple \href{https://doi.org/10.1103/PhysRevLett.77.3865}{Phys. Rev. Lett. \textbf{77}, 3865 (1996).}
\bibitem{Liechtenstein:prb} A. I. Liechtenstein, V. I. Anisimov, and J. Zaanen, Density-functional theory and strong interactions: Orbital ordering in Mott-Hubbard insulators \href{https://doi.org/10.1103/PhysRevB.52.R5467}{Phys. Rev. B \textbf{52}, R5467 (1995).}
\bibitem{Mostofi:cpc} A. A. Mostofi, J. R. Yates, Y. S. Lee, I. Souza, D. Vanderbilt, and N. Marzari, wannier90: A tool for obtaining maximally-localised Wannier functions \href{https://doi.org/10.1016/j.cpc.2007.11.016}{Comput. Phys. Commun. \textbf{178}, 685 (2007).}
\bibitem{Graser2009} S. Graser, T. A. Maier, P. J. Hirschfeld, and D. J.  Scalapino, Near-degeneracy of several pairing channels in multiorbital models for the Fe pnictides \href{https://doi.org/10.1088/1367-2630/11/2/025016}{New J. Phys. \textbf{11}, 25016 (2009).}
\bibitem{Kubo2007} K. Kubo, Pairing symmetry in a two-orbital Hubbard model on a square lattice \href{https://doi.org/10.1103/PhysRevB.75.224509}{Phys. Rev. B \textbf{75}, 224509 (2007).}
\bibitem{Romer2020} A. T. R{\o}mer, T. A. Maier, A. Kreisel, I. Eremin, P. J. Hirschfeld, and B. M. Andersen, Pairing in the two-dimensional Hubbard model from weak to strong coupling \href{https://doi.org/10.1103/PhysRevResearch.2.013108}{Phys. Rev. Res. \textbf{2}, 013108 (2020).}
\bibitem{Altmeyer2016} M. Altmeyer, D. Guterding, P. J. Hirschfeld, T. A. Maier, R. Valent\'{\i}, and D. J. Scalapino, Role of vertex corrections in the matrix formulation of the random phase approximation for the multiorbital Hubbard model \href{https://doi.org/10.1103/PhysRevB.94.214515}{Phys. Rev. B \textbf{94}, 214515 (2016).}
\bibitem{Mishra:sr} V. Mishra, T. A. Maier and D. J. Scalapino, $s^{\pm}$ pairing near a Lifshitz transition \href{https://doi.org/10.1038/srep32078}{Sci Rep. \textbf{6}, 32078 (2016).}
\bibitem{Maier:prb22} T. A. Maier, and Elbio Dagotto, Coupled Hubbard ladders at weak coupling: Pairing and spin excitations \href{https://doi.org/10.1103/PhysRevB.105.054512}{Phys. Rev. B \textbf{105}, 054512 (2022).}
\bibitem{Supplemental}{See Supplemental Material for method details, wannier fittings, more tight-binding results, additional discussion for the RPA calculations, and thicker films.}
\bibitem{Momma:vesta} K. Momma and F. Izumi, VESTA 3 for three-dimensional visualization of crystal, volumetric, and morphology data \href{https://doi.org/10.1107/S0021889811038970}{J. Appl. Crystallogr. \textbf{44}, 1272 (2011).}
\bibitem{RPA-U} {Since a spin-density wave instability was obtained at $U  =  0.8$ eV at $n = 3.0$, we used $U  =  0.7$ eV in the subsequent RPA calculations to maintain the continuity of the $\lambda$ vs filling $n$.}
\bibitem{Lin:prl21} L. F. Lin, Y. Zhang, G. Alvarez, A. Moreo, and E. Dagotto, Origin of Insulating Ferromagnetism in Iron Oxychalcogenide Ce$_2$O$_2$FeSe$_2$, \href{https://doi.org/10.1103/PhysRevLett.127.077204}{Phys. Rev. Lett. \textbf{127}, 077204 (2021).}
\bibitem{Lin:cp} L. F. Lin, Y. Zhang, G. Alvarez, M, A, McGuire, A. F. May, A. Moreo, and E. Dagotto, Stability of the interorbital-hopping mechanism for ferromagnetism in multi-orbital Hubbard models \href{https://doi.org/10.1038/s42005-023-01314-w}{Commun. Phys. \textbf{6}, 199 (2023).}
\bibitem{Lin:prb22} L. F. Lin, Y. Zhang, G. Alvarez, A. Moreo, Jacek Herbrych, and E. Dagotto, Prediction of orbital-selective Mott phases and block magnetic states in the quasi-one-dimensional iron chain Ce$_2$O$_2$FeSe$_2$ under hole and electron doping\href{https://doi.org/10.1103/PhysRevB.105.075119}{Phys. Rev. B \textbf{105}, 075119 (2022).}
\end{references}
\end{document}